\documentclass {vgtc}                          

\graphicspath{{figures/}{pictures/}{images/}{./}} 

\usepackage{times}                     

\usepackage{tabu}                      
\usepackage{booktabs}                  
\usepackage{lipsum}                    
\usepackage{mwe}                       

\usepackage{mathptmx}                  

\usepackage[hyphens]{url}
\usepackage{cite}
\onlineid{0}

\vgtccategory{Research}

\vgtcinsertpkg

\title{Connections Beyond Data: Exploring Homophily With Visualizations}

\author{Poorna Talkad Sukumar\thanks{e-mail: pt2393@nyu.edu} 
\qquad Maurizio Porfiri\thanks{e-mail: mporfiri@nyu.edu}
\qquad Oded Nov\thanks{e-mail: onov@nyu.edu}\\ %
     \scriptsize New York University }

\teaser{
  \centering
  \includegraphics[width=\linewidth]{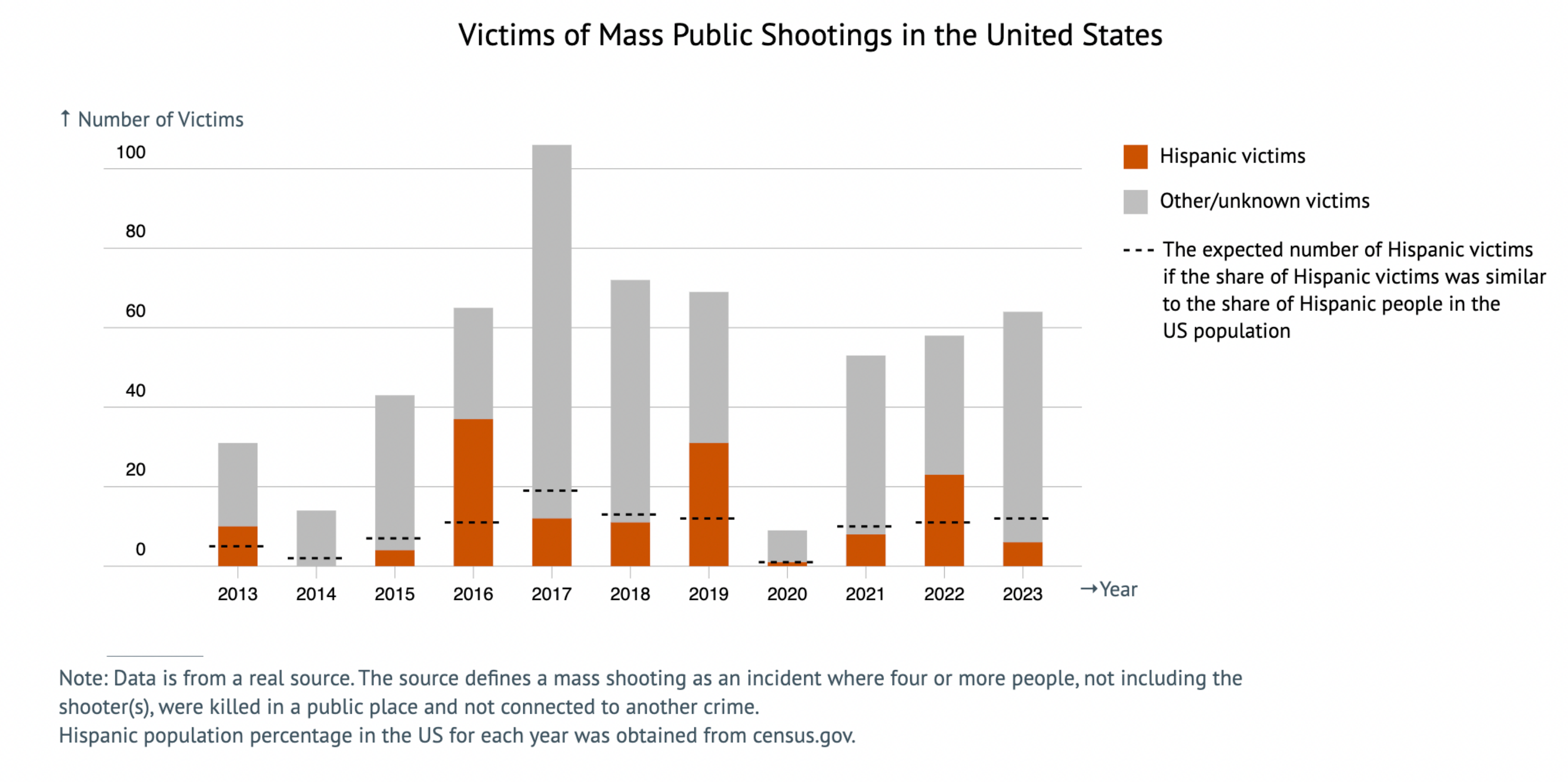}
  
  \caption{One of the three conditions used in our experiment consisting of a bar chart of the counts of victims in mass shootings in the United States from 2013 to 2023, highlighting the counts of Hispanic victims. The other two conditions consist of the same bar chart but highlight the counts of White and Black victims, respectively. }
  \label{fig:teaser}
}

\abstract{
    Homophily refers to the tendency of individuals to associate with others who are similar to them in characteristics, such as, race, ethnicity, age, gender, or interests. In this paper, we investigate if individuals exhibit racial homophily when viewing visualizations, using mass shooting data in the United States as the example topic. We conducted a crowdsourced experiment (N=450) where each participant was shown a visualization displaying the counts of mass shooting victims, highlighting the counts for one of three racial groups (White, Black, or Hispanic). Participants were assigned to view visualizations highlighting their own race or a different race to assess the influence of racial concordance on changes in affect (emotion) and attitude towards gun control. 
    While we did not find evidence of homophily, the results showed a significant negative shift in affect across all visualization conditions. Notably, political ideology significantly impacted changes in affect, with more liberal views correlating with a more negative affect change. Our findings underscore the complexity of reactions to mass shooting visualizations and suggest that future research should consider various methodological improvements to better assess homophily effects.

} 

\CCScatlist{
Visualization; Journalism; Mass shootings; Race; Homophily
}

\begin{document}


\firstsection{Introduction}

\maketitle

Homophily is the tendency for individuals to associate with others who are similar to themselves in various attributes, including age, gender, race, education, occupation, and beliefs \cite{mcpherson2001birds}. Visualization research has primarily focused on ways to visualize homophily, for example, in node-link diagrams \cite{reimann2022color}.
However, there is a notable gap in the literature concerning how people exhibit homophily when they view and interpret visualizations that might prompt them to identify and associate with certain aspects of the presented data.  
In this paper, we aim to bridge this gap by investigating whether individuals exhibit racial homophily by using visualizations of mass shooting data in the United States.

Recent research indicates that the race of victims in mass shootings affects public opinion and subsequently influences gun policy responses; specifically, mass shootings with predominantly White victims lead to more restrictive firearm laws compared to those with primarily minority victims \cite{markarian2024racially}. Considering that people often hold strong attitudes on this issue of mass shootings and gun control, we focus on the race of victims of mass shootings to explore homophily in visualizations. In particular, we aim to understand if people are more emotionally affected and if their views on gun control change when victims of mass shootings belong to their own racial group. 

We conducted a pre-registered\footnote{\url{https://osf.io/mnc8s/?view_only=74d37cc7425744eab79d6a5180bb3a08}} crowdsourced experiment where participants were each shown one of three visualization conditions, each displaying the number of victims from mass shootings and highlighting the counts for a specific racial group: White, Black, or Hispanic (see Figure \ref{fig:teaser}). For each condition, we assigned both participants who belonged to the racial group highlighted in the visualization, and participants who were from a different racial group. This setup allowed us to assess how racial concordance influences changes in affect and attitude towards gun control.



Our research provides an initial exploration of homophily in visualizations, specifically focusing on race, and makes the following contributions: (1) we present empirical evidence from a crowdsourced experiment demonstrating the absence of racial homophily in viewers' emotional and attitudinal responses to mass shooting visualizations; (2) we identify political ideology as a significant factor influencing affective responses, with more liberal views correlating with a more negative change in affect; and (3) we highlight the need for future studies to investigate alternative visualization designs, explore different societal issues, and use larger sample sizes with rigorous statistical analyses to detect subtle homophily effects.


\section{Background}


Homophily, the principle that similar individuals tend to associate with each other more than with dissimilar ones, manifests in various forms across social contexts. McPherson et al. \cite{mcpherson2001birds} have provided seminal insights into this phenomenon, emphasizing its pervasive impact on social networks. Their research demonstrates that people often form connections based on shared attributes such as race, ethnicity, age, and interests, making homophily a fundamental element in the analysis of social structures. They define two primary types of homophily: \textit{status} homophily, which is based on observable social attributes, such as, race or occupation, and \textit{value} homophily, which centers on shared beliefs, attitudes, and values. Among various forms of homophily, racial homophily stands out as particularly consequential. It can both exacerbate social divisions and inequalities while also serving to strengthen support networks within marginalized communities \cite{mollica2003racial}.

Understanding homophily is important for shaping effective policies and strategies across various domains because it reveals the inherent patterns of connection within social networks. 
Recognizing patterns of homophily can assist in designing better communication strategies, creating more inclusive environments, and managing diversity effectively within both social and professional settings \cite{mcpherson2001birds}. 
For example, homophily can inform targeted communication strategies by aligning messages with specific preferences and values of audiences, thereby improving effectiveness in marketing, health campaigns, and political messaging. 

In human-computer interaction (HCI), homophily has been explored in the context of online communities and social interactions. Studies show strong preferences for potential partners with similar characteristics on online dating platforms \cite{fiore2005homophily}, and diverse roles of network connections in influencing users’ actions on social media \cite{brzozowski2008friends}. These findings underline the profound influence of homophily on online interactions and community dynamics, with significant implications for the design and management of digital platforms.

In visualization, the focus shifts to how these concepts of homophily can be visualized. For example, Reimann et al. \cite{reimann2022color} found that color-coded links in node-link diagrams significantly enhance the perception of homophily, demonstrating that such visual enhancements in graph designs can improve the understanding of complex social relationships. Visualization research has also explored how individual differences \cite{ziemkiewicz2012understanding, liu2020survey}, such as personality traits, as well as pre-existing attitudes and beliefs \cite{kim2017explaining, xiong2022seeing} affect visualization interaction and interpretation. Studying homophily with visualizations also involves recognizing individual differences, particularly social, demographic, or ideological characteristics, but specifically examines how visualizations can resonate differently with individuals, owing to these differences.


\section{Experiment}

We set up our study as an online survey on Qualtrics and we recruited participants through the Prolific platform. We set out to answer the following research questions with our study.\\

When viewing a visualization of victim counts in mass shootings highlighting counts belonging to a specific race: \\
RQ1. How does concordance between the viewer's race and the highlighted race influence changes in their emotional responses? \\
RQ2. How does  concordance between the viewer's race and the highlighted race influence changes in their attitudes towards gun control? 

For RQ1 and RQ2, we expected significant differences in affect and attitude changes between participants of the same racial group and those from different racial groups. Specifically, we anticipated more negative changes in affect and an attitude shift towards stricter gun control among participants of the same racial group.

\subsection{Participants}
We recruited 450 participants through the Prolific platform. Participants were required to be at least 18 years old, be fluent in English, be located in the United States, and complete the survey on a desktop or laptop device. All the survey responses were de-identified. We discarded responses that failed a verification question. The median completion time across all experimental conditions was approximately seven minutes, and the participants were paid \$2 for their participation.

\begin{table*}[!t]
\centering
\caption{The number of participants who indicated a change in valence (emotional response) across different visualization conditions and participant assignments encoded with a heatmap representation where lighter colors correspond to lower values and vice versa. Negative values represent  negative changes in emotion and vice versa. }
  \includegraphics[width=1\textwidth]{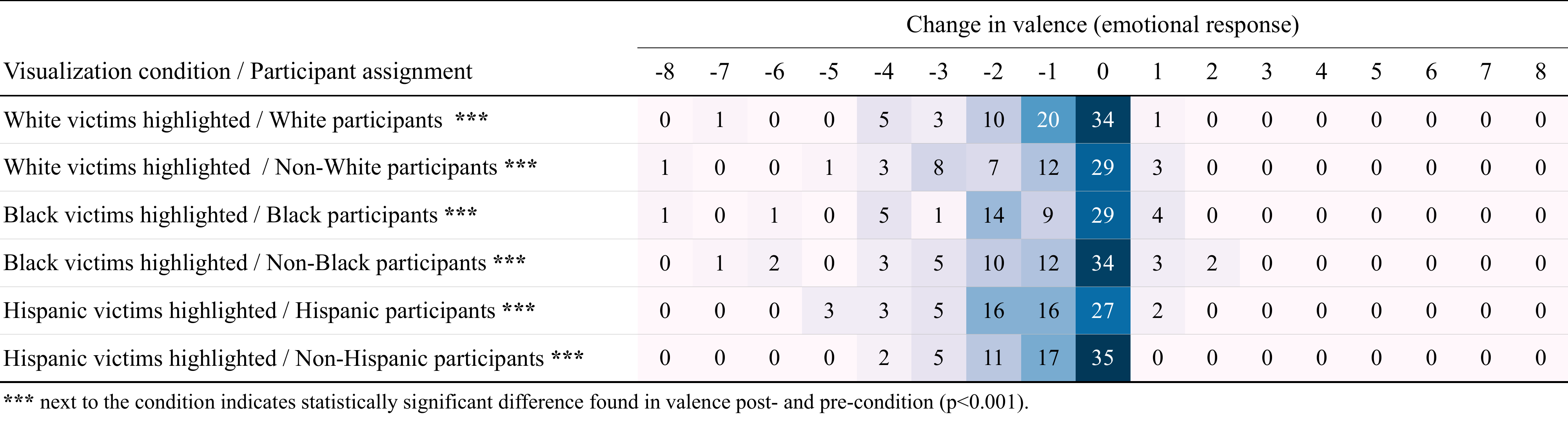}
   \label{tab1}
\end{table*}

\begin{table}[!t]
\centering
\caption{The number of participants who indicated a change in attitude towards gun control. Positive values represent changes in attitude towards stricter gun laws and vice versa.}
  \includegraphics[width=0.47\textwidth]{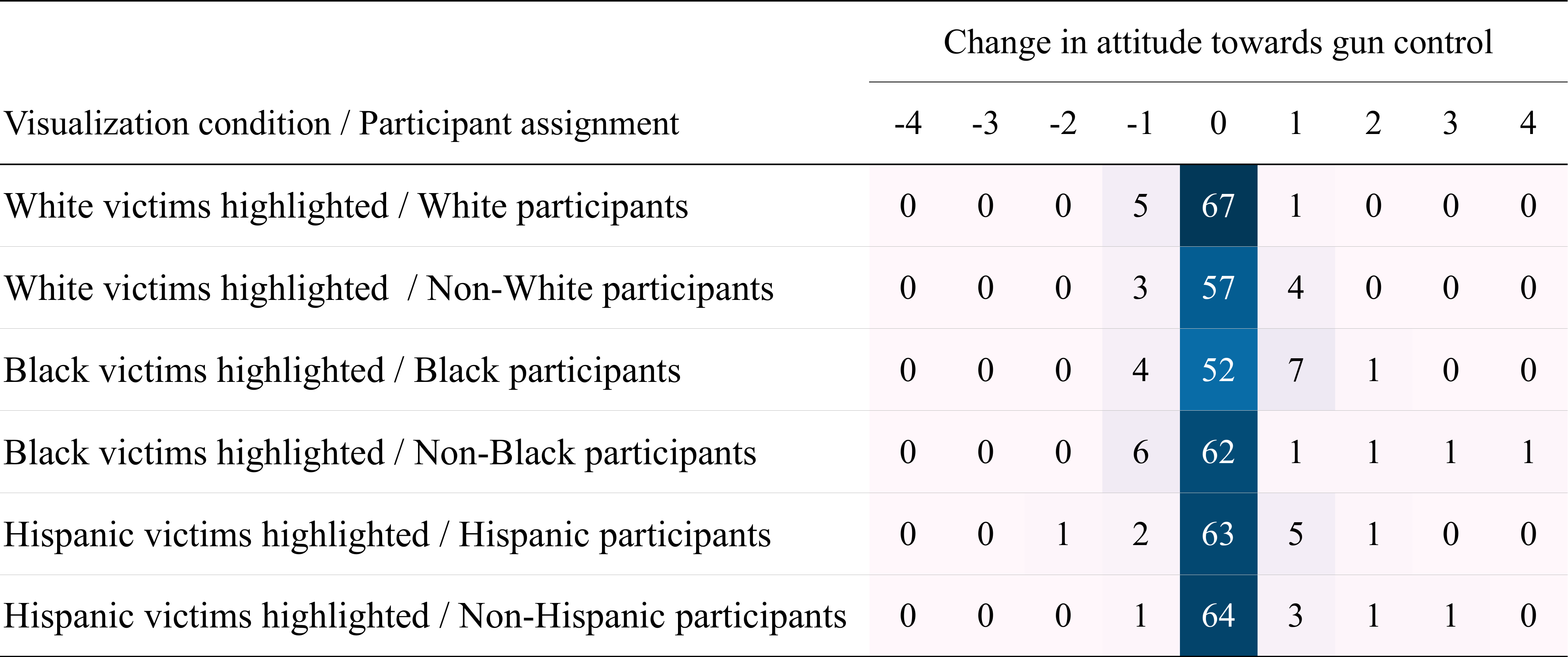}
   \label{tab2}
\end{table}

\subsection{Study Material}

We obtained data for our visualization conditions from the Violence Project Mass Shooter Database \cite{violenceproject}, which, to our knowledge, is the only public database that includes the race of mass shooting victims in the U.S. We used bar charts to display the victim counts, as bar charts are commonly used by the media for presenting this type of data. We focused on the period from 2013 to 2023 to present both recent and sufficiently longitudinal data. We designed three experimental conditions, each featuring a bar chart that showed victim counts and also specifically highlighted the victim counts of one of the three racial groups: White, Black, or Hispanic. We chose to highlight these three racial groups because they had the highest victim counts in the dataset. 

The number of victims varies across the three racial groups. To provide adequate context for interpretation, we not only highlighted the actual victim counts but also included the expected number of victims for each group, based on their proportion in the U.S. population (see Figure \ref{fig:teaser}). This calculation uses annual U.S. population data by race/ethnicity from the Census Bureau\footnote{\url{census.gov}}. The disparities between the actual and expected victim counts for the three racial groups are perceptually similar, with no pronounced differences for any group, thereby ensuring comparability across all the experimental conditions. All the conditions were static and are included in the supplementary material.

\subsection{Survey Procedure}

For each of the three visualization conditions, we assigned 75 participants who identified with the same racial group highlighted in the visualization, and 75 participants who identified with a different racial group. We facilitated these assignments using the prescreening options on Prolific, enabling us to filter participants according to their self-reported race/ethnicity.

The experiment consisted of three stages. In Stage 1, participants were asked to fill out a demographic survey on their gender, age, ethnicity, and education level. They were also asked about their views on gun control, specifically whether they believed firearm laws should be made stricter. This was measured using a 5-point Likert scale, with options ranging from ``Strongly disagree" to ``Strongly agree". Then, participants' emotional responses were assessed using the 9-point Self-Assessment Manikin (SAM) scale for measuring valence \cite{lang2005international}, a method that has been effectively used to study affective priming in visualizations \cite{harrison2013influencing}.

In Stage 2, participants were shown their assigned visualization condition. Similar to the study by Kong et al. \cite{kong2018frames}, participants were required to view the visualization condition for at least 30 seconds before the option to proceed appeared. The same visualization condition was shown on the next page and we asked two factual questions below based on its content. In the following page, we included a verification question that asked for the topic of the visualization. 

Stage 3 began with participants once again rating their emotional response using the 9-point SAM scale for measuring valence, the same as in Stage 1. They were then asked to reevaluate their views on gun control, identical to the initial query, and to provide reasons for their response. We then asked participants to write about what they learned from the visualization condition in detail. Questions regarding their political partisanship, frequency of news consumption, and familiarity with visualizations followed. Finally, participants were given a short debrief explaining the purpose of the experiment. The Qualtrics survey is included as part of the supplementary material.

\section{Results}

Each participant was assigned a single visualization condition. We excluded participants who incorrectly answered the topic verification question, as well as those that failed to answer both the factual questions. In total, 35 of the 450 participants were removed. Each of the three conditions has an average of 70 participants from the same racial group and an average of 69 participants from a different racial group.

\subsection{RQ1. Influence of racial concordance on changes in emotional responses}
\label{rq1}


We calculated the change in valence by subtracting the pre-condition valence from the post-condition valence. To examine the impact of racial concordance on affect changes, we compared changes in valence from participants of the same racial group with those of different racial groups across all conditions using a non-parametric Mann Whitney \textit{U} test. 
We found no statistically significant differences between the two groups (Z=-0.87, p=0.39).


We conducted a non-parametric Wilcoxon signed-rank test separately for each condition and participant assignment (same race and different race), comparing pre- and post-condition valence. Significant differences were found in all conditions and participant assignments (p$<$0.001), with post-condition valence consistently lower than pre-condition valence (see Table \ref{tab1}).

No statistically significant differences were observed between participant assignments (same race vs. different race) for each condition for changes in valence using the Mann Whitney \textit{U} test. Similarly, a Kruskal-Wallis test showed no significant differences in valence change across all three conditions, regardless of participant assignment.

\subsection{RQ2. Influence of racial concordance on changes in attitude}
\label{rq2}

We calculated the change in attitude by subtracting the pre-condition attitude from the post-condition attitude. To assess the influence of racial concordance on attitude changes, we conducted the same statistical tests as those described in Section \ref{rq1} on changes in attitude instead of valence. However, we observed minimal attitude changes overall, and, as a result, we did not find statistically significant differences in any of the tests (see Table \ref{tab2}).




\subsection{Additional Analysis}
\label{rq4}

We performed a multiple linear regression to examine the effects of age, gender, educational background, political views, and racial concordance (same race or different race) on the changes in valence. The results showed that the only statistically significant predictor was political views (t = -4.079, p$<$0.001), indicating that more liberal political views are associated with more negative changes in valence.

The model’s fit was relatively low, with an R-squared value of 4.747\% and an adjusted R-squared of 3.318\%, suggesting that only a small proportion of the variance in the change in valence is explained by the predictors included in the model. The F-statistic for the overall model was significant (F(6, 400) = 3.322, p$<$0.01), indicating that the model was statistically significant at explaining some of the variance in valence changes.

\section{Discussion}

\subsection{Interpreting the Absence of Homophily }

Our study sought to explore the presence of racial homophily in the context of mass shooting visualizations, a setting where emotional and cognitive reactions are particularly salient. Contrary to our expectations, our findings did not demonstrate a significant effect of racial concordance on either emotional responses or attitudes towards gun control. This absence of observed homophily could suggest that the stark reality of mass shootings, as presented through data visualizations, may override the typical in-group favoritism that homophily predicts. It is possible that the impact of the visualized data is so universally negative that it saturates emotional responses across different racial viewpoints, homogenizing reactions rather than differentiating them.

Prior literature on visualization's impact on affect and prosociality suggests that the effects, if any, are often subtle and require large sample sizes to detect reliably. For instance, Morais et al. reviewed and conducted large-sample studies on anthropographics, finding that their effects on prosocial behavior are minimal and difficult to measure without substantial statistical power\cite{morais2021can}. 
Future studies on homophily could similarly consider larger sample sizes and more rigorous power analyses to detect these subtle effects, ensuring that the estimated effect sizes are adequately accounted for and that the studies are sufficiently powered to uncover small but potentially meaningful impacts.

Previous research \cite{kong2018frames, heyer2020pushing} has also underscored the difficulty of effecting attitude changes through visual stimuli alone, especially with such controversial topics, where opinions may be more entrenched.  
Heyer et al. emphasized the importance of eliciting prior knowledge and beliefs to understand how visualizations affect attitudes and perceptions, especially on contentious topics. Their findings indicated that while eliciting prior beliefs can increase engagement and surprise, it does not necessarily translate to significant attitude change \cite{heyer2020pushing}. Additionally, Kong et al. highlighted the role of framing in visualization titles and its potential to bias viewers' interpretation and recall of information. While their study found that titles could influence perceived messages, they did not significantly affect overall attitude change \cite{kong2018frames}.  This underscores the complexity of influencing deeply held beliefs and attitudes through visual data alone, particularly in politically charged contexts. 
It is possible that our findings could diverge if we examined less politically charged and controversial topics. 

\subsection{The Role of Political Ideology}

Interestingly, our results revealed that political ideology plays a more definitive role in shaping reactions to the mass shooting visualizations. Participants with more liberal ideologies exhibited significantly more negative affective responses, aligning with previous research indicating that liberals are more likely to support stricter gun control measures \cite{oraka2019cross}. 

The influence of political ideology observed in our study is consistent with prior research indicating that personal beliefs and prior attitudes influence the interpretation of visual data \cite{markant2023data, kim2017explaining, heyer2020pushing, xiong2022seeing, ottley2015manipulating, talkad2024are, pandey2014persuasive}. For example, Pandey et al. demonstrated that visualizations could be persuasive, but their effectiveness is often moderated by pre-existing attitudes and beliefs \cite{pandey2014persuasive}. 
This implies that, for contentious issues influenced by existing political orientations, implementing more detailed controls and soliciting finer-grain prior beliefs may be necessary to accurately identify any effects.

\subsection{Implications for Visualization Design}

Our study shows that even relatively simple visualizations can evoke strong negative emotions, suggesting that complex design features or dramatic embellishments, such as those used in the `Gun Deaths' visualizations with the inverted \textit{y}-axis \cite{florida} or by Periscopic \cite{periscopic}, may not always be necessary to engage viewers emotionally. This finding makes the work by Lee-Robbins et al. particularly relevant, which advocates for the explicit articulation of affective intents in visualizations, especially when dealing with politically-charged content \cite{lee2022affective}.  
The significant emotional response elicited by our visualizations underscores the importance of intentionality in design; specifically, it highlights the need to clearly communicate intended emotional outcomes, whether to encourage concern for the topic, reinforce or change attitudes, or prompt actions. By adopting a structured approach to defining these affective intents, designers can more effectively ensure that the emotional impact of their visualizations aligns with their communicative objectives.
For example, an objective for mass shooting visualizations could be a call to action, such as advocating for policy changes.


\subsection{Future Research Directions}

Given the absence of homophily effects in our study, further research is necessary to understand under what conditions, if any, homophily might emerge in the interpretation of visual data. Future studies should explore alternative visualization designs. For example, prior work shows that bar charts can be overly simplistic for presenting social outcomes \cite{holder2022dispersion}. They obscure within-group variability, which can promote deficit thinking by attributing disparities to the individuals being visualized rather than to external factors, potentially leading to stereotyping \cite{holder2022dispersion, jonathan2021no}. Therefore, future research could investigate visualization designs that show within-group variability, such as distributions across race or charts that present individual victims' data, rather than aggregate sums. Additionally, integrating insights from anthropographics literature could be valuable, as these approaches aim to enhance empathy and prosocial behavior by incorporating human elements into visualization design \cite{elli2020tied, morais2021can, dhawka2023we, boy2017showing}.

Future studies could also explore different societal issues, such as public health campaigns or gender equality initiatives, where viewer characteristics might interact more noticeably with content. Examining other demographic variables such as age or gender might reveal different patterns of homophily or provide further insight into the complex interplay of viewer characteristics and data interpretation.

Our findings also indicate that affect alone may not be a sufficient measure of homophily in the context of visual data interpretation. Future research could consider alternative metrics, such as, memorability, engagement, or enjoyment, or the quality of viewers' reasoning as they interact with data visualizations. Such measures may provide deeper insights into how visualizations influence interpretation based on viewer characteristics. 


\section{Conclusion}


Our study aimed to explore racial homophily in visualizations of mass shooting data and its influence on viewers' emotions and attitudes towards gun control. Contrary to expectations, we found no significant evidence of racial homophily affecting these responses; instead, we found a negative shift in affect across all groups which suggested that the severity of the topic might have overshadowed individual racial affinities. Interestingly, political ideology was a more significant factor, with participants holding more liberal views experiencing more negative response changes, indicating that ideological orientation may influence the interpretation of visual data more than demographic similarities. Further research is required to identify any scenarios where homophily might influence the interpretation of visual data. Future studies should consider different societal issues and types of data visualizations, 
and also employ larger sample sizes and additional controls to detect subtle effects.

\section{Supplementary Materials}

Our supplementary materials include the following: the Qualtrics survey, visualization conditions, de-identified survey data, a revision letter addressing reviewer comments, ethical considerations, and additional figures. The materials can be found at \url{https://osf.io/3crqx/}.

\acknowledgments{
This work was supported by the National Science Foundation under Grant CMMI-1953135.}

\Urlmuskip=0mu plus 1mu\relax

\bibliographystyle{abbrv-doi}

\bibliography{template}
\end{document}